\documentclass[conference]{IEEEtran}
\IEEEoverridecommandlockouts

\usepackage{cite}
\usepackage{amsmath,amssymb,amsfonts}
\usepackage{algorithmic}
\usepackage{graphicx}
\usepackage{textcomp}
\usepackage{xcolor}
\usepackage{url}

\makeatletter

\DeclareRobustCommand*{\authorrefmark}[1]{\raisebox{0pt}[0pt][0pt]{\textsuperscript{#1}}}%

\makeatother

\def\BibTeX{{\rm B\kern-.05em{\sc i\kern-.025em b}\kern-.08em
    T\kern-.1667em\lower.7ex\hbox{E}\kern-.125emX}}

\begin{document}

\title{Nonlinear Feedback in Josephson Circuit Optimization: Application to a Kerr-Reversal JTWPA
\thanks{This work is partially supported by the European project MiSS. MiSS is funded by the European Union through the Horizon Europe 2021-2027 Framework Programme, Grant agreement ID: 101135868. This work is partially supported by the European project MetSuperQ. The 23FUN08 MetSuperQ project has received funding from the European Partnership on Metrology, co-financed from the European Union’s Horizon Europe Research and Innovation Programme and by the Participating States. This work is partially supported by the European project QuTest. Qu-Test has received funding from the European Union’s Horizon Europe – The EU research and innovation programme under the Grant Agreement numbers 101113901.}
}

\author{%
\IEEEauthorblockN{%
Emanuele Palumbo\authorrefmark{\#\$1},
Alessandro Alocco\authorrefmark{\#\$2},
Andrea Celotto\authorrefmark{\#\$3},
Luca Fasolo\authorrefmark{\$4},
Bernardo Galvano\authorrefmark{*\$5},\\
Patrizia Livreri\authorrefmark{*6},
Emanuele Enrico\authorrefmark{\$7}
}
\IEEEauthorblockA{%
\authorrefmark{\#}Department of Applied Science and Technology, Politecnico di Torino, Torino, Italy\\
\authorrefmark{\$}Quantum Metrology and Nanotechnology Division, Istituto Nazionale di Ricerca Metrologica, Torino, Italy\\
\authorrefmark{*}Department of Engineering, University of Palermo, Palermo, Italy\\
\authorrefmark{7}e.enrico@inrim.it
}
}
\maketitle

\begin{abstract}
Optimizing Josephson-based nonlinear microwave devices is computationally demanding because the straightforward approach requires exploring broad circuit parameter spaces through expensive nonlinear simulations. Josephson Circuits Optimizer addresses this problem by using harmonic
balance simulations in two stages: fast linear simulations to select
promising circuit configurations according to properties such as
impedance and phase matching, followed by nonlinear simulations to
optimize their operating conditions. However, pump-induced effects such as impedance renormalization and Kerr-induced modifications of phase matching are not captured during the initial linear stage.
We therefore introduce a nonlinear feedback extension that transfers information from the pumped response back to the linear optimization.

The approach is investigated using a Josephson traveling-wave parametric amplifier with a reversed-Kerr architecture, whose model is validated against experimental gain measurements. 
We then evaluate the difference between the linear and pumped response values of a metric based on the input reflection coefficient, while excluding configurations with insufficient third-harmonic suppression. 
This discrepancy is considered a candidate feedback observable, as it exhibits features similar to the gain landscape, supporting its use in subsequent optimization cycles.
\end{abstract}


\section{Introduction}

Superconducting circuits based on Josephson junctions (JJs) enable key functions for quantum information processing and microwave quantum engineering, including quantum-limited amplification, frequency conversion, squeezed-state generation, and high-fidelity qubit readout \cite{fasolo_review,twpa_perspective,macklin_jtwpa}.

For distributed Josephson devices such as Josephson Traveling-Wave
Parametric Amplifiers (JTWPAs), the large number of circuit elements, strong nonlinearity, and high-dimensional design spaces make optimization computationally demanding.
Although harmonic balance methods provide an efficient steady-state solution compared to time-domain approaches \cite{peng_xparameters,shiri_hb}, exploring broad device parameter spaces remains computationally expensive.

To address this challenge, we previously introduced
Josephson Circuits Optimizer (JCO) \cite{jco,jco_github}, a simulation framework
built on JosephsonCircuits.jl \cite{josephsoncircuits}, designed to
optimize device performance, such as gain, while limiting
computational cost.
JCO uses harmonic balance simulations in two main stages.
First, computationally efficient linear simulations identify promising
circuit configurations using properties such as impedance and phase
matching \cite{obrien_phase_matching}.
Second, nonlinear simulations are performed on the selected
configuration under pump excitation to optimize its operating
conditions and evaluate its performance.

However, this single-cycle approach is limited because the linear stage does not account for nonlinear effects emerging under strong pumping, such as impedance renormalization and Kerr-induced modifications of the dispersion relation \cite{kern_reflections,peng_floquet,ranadive_kerr,bell_inverse_kerr}.
Consequently, the configuration identified as optimal by the linear model may no longer remain optimal under pumped operation. 
We therefore introduce a nonlinear feedback extension of the original JCO workflow, in which information from the pumped response is fed back into the linear stage of the subsequent optimization cycle.

To investigate this feedback strategy, we consider a four-wave-mixing (4WM) JTWPA from Silent Waves with a reversed-Kerr architecture \cite{ranadive_kerr}. Its nonlinear model is first validated against gain measurements at different pump frequencies and powers. The variation of the reflection coefficient between the linear and pumped responses is then investigated as a candidate feedback observable, while excluding configurations with insufficient third-harmonic suppression.


\section{Method}
\label{sec:method}

\begin{figure}[!t]
\centerline{\includegraphics{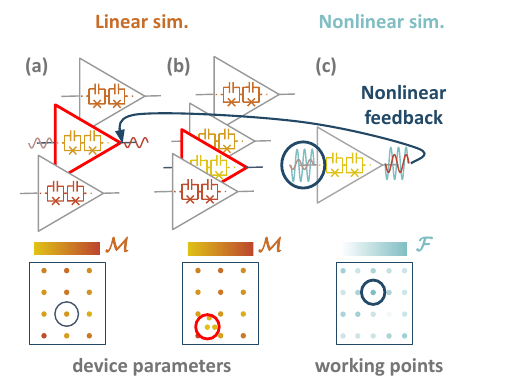}}
\caption{Overview of the JCO optimization workflow.
(a) The device parameter space is explored through linear simulations
using the metric $\mathcal{M}$.
(b) Linear optimization identifies the selected circuit configuration.
(c) The selected configuration is analyzed under pumped conditions,
and information from the pumped response is fed back into the linear
stage of the subsequent optimization cycle.}
\label{fig:JCO_WF}
\end{figure}

The original JCO framework separates the optimization of the circuit configuration from the optimization of its operating conditions, as illustrated in Fig.~\ref{fig:JCO_WF}.

First, the device parameter space $P$, defined by the values of circuit parameters such as capacitances, inductances, and critical currents, is explored through computationally efficient linear simulations. Each circuit configuration $p\in P$ is characterized by its scattering parameters $S(p)$ and evaluated through a device-specific metric $\mathcal{M}$, yielding the optimal circuit configuration
\begin{equation}
p^*=\arg\min_{p\in P} \mathcal{M}(S(p)).
\end{equation}

The selected configuration $p^*$ is then analyzed through nonlinear harmonic balance simulations over the working-point space $Q$, defined by the frequencies and powers of the input signals. A performance function $\mathcal{F}$, such as the average gain, determines the optimal operating point:
\begin{equation}
q^*=\arg\max_{q\in Q} \mathcal{F}(X(q;p^*)),
\end{equation}
where $X$ denotes the nonlinear X-parameters of the circuit.
Because the linear optimization metric may no longer remain
representative under strong pumping, information extracted from the
nonlinear response is used to update it before the next iteration.

The nonlinear correction is represented by the correction term $\mathcal{C}$. Starting from an initial metric $\mathcal{M}^{(0)}$, the original JCO workflow identifies the optimal circuit configuration $p^{(0)}$ and the corresponding working point $q^{(0)}$. The nonlinear response is then used to construct $\mathcal{C}^{(0)}$, yielding the updated metric
\begin{equation}
\mathcal{M}^{(n+1)}
=
g\!\left(\mathcal{M}^{(n)},\mathcal{C}^{(n)}\right),
\end{equation}
where $g$ defines how the nonlinear correction is incorporated into the linear metric function.
The updated metric defines the next optimization target,
\begin{equation}
p^{(n+1)}
=
\arg\min_{p\in P} \mathcal{M}^{(n+1)}(S(p)).
\end{equation}

The procedure can be repeated until successive corrections become
negligible. The feedback infrastructure is implemented within JCO,
while the present work investigates the observable used to construct
the correction operator $\mathcal{C}$.

\section{Application to the Kerr reversal Josephson Traveling-Wave Parametric Amplifier}
\label{sec:Application}

\subsection{Device model and experimental validation}

\begin{figure}[!t]
\centerline{\includegraphics{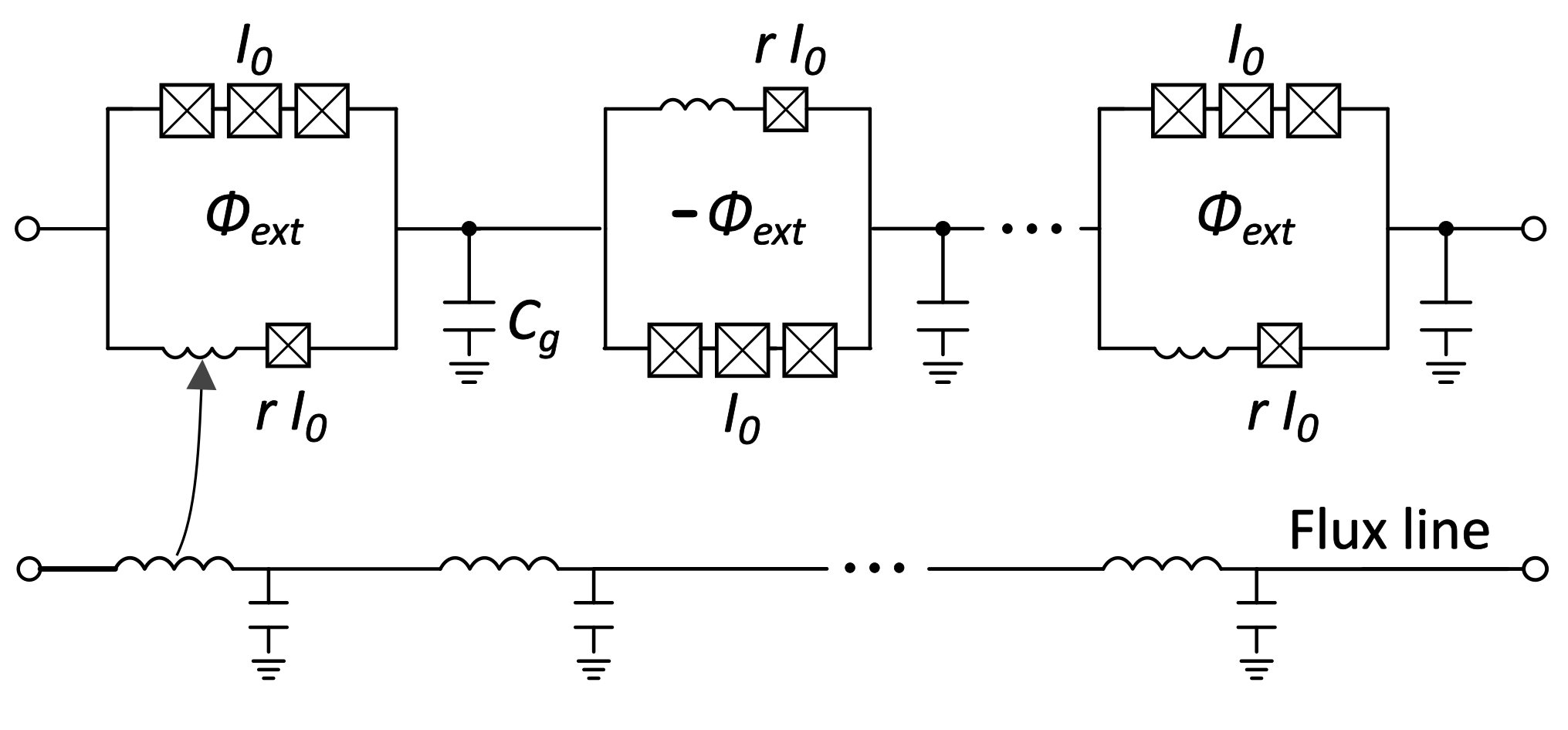}}
\caption{Equivalent lumped-element circuit of the 4WM JTWPA adopted
in this work, following the model reported in \cite{ranadive_kerr}.
The transmission line is formed by a periodic array of unit cells,
each comprising a Superconducting Nonlinear Asymmetric Inductive
eLement (SNAIL) and a ground capacitance $C_g$. Each SNAIL consists
of one small JJ with critical current $rI_0$ and three large JJs with
critical current $I_0$, with $r=0.07$ the junction asymmetry function.
The magnetic bias is represented by a periodic flux line mutually
coupled to the SNAIL inductive branches, producing the external flux
$\Phi_{\mathrm{ext}}$. Adjacent cells have reversed orientations,
resulting in alternating flux biases $\pm\Phi_{\mathrm{ext}}$ and
suppressing the effective three-wave-mixing contribution while
preserving the four-wave-mixing response. In the experimental device, the
corresponding magnetic bias is generated by a superconducting coil.}
\label{fig:scheme}
\end{figure}

The proposed framework is evaluated using the lumped-element model of a 4WM JTWPA from Silent Waves \cite{ranadive_kerr}. The corresponding equivalent circuit, reproducing the reversed-Kerr architecture, is shown in Fig.~\ref{fig:scheme}.

\begin{figure}[!t]
\centerline{\includegraphics{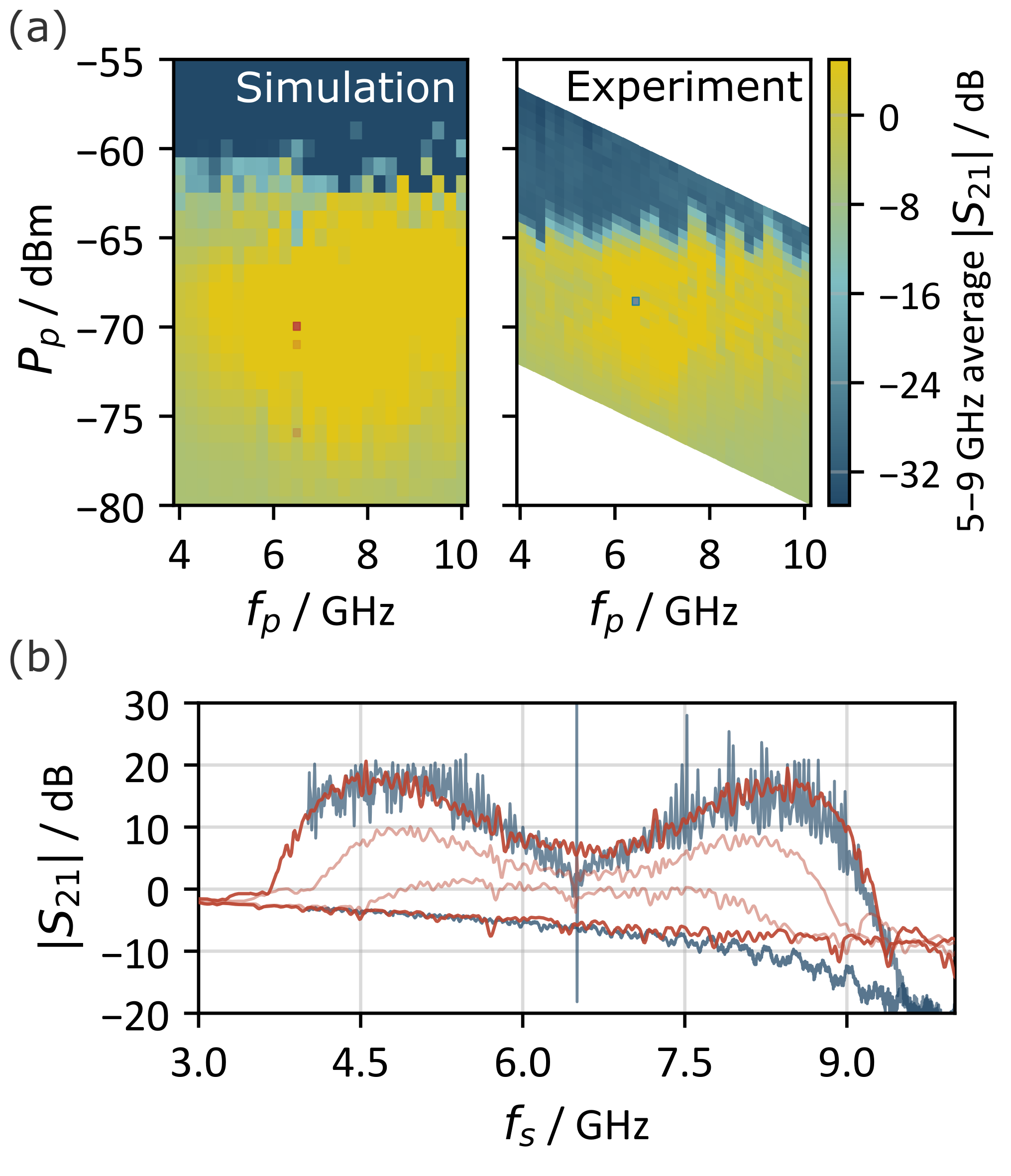}}
\caption{Comparison between simulated and measured amplifier performance 
for the same device configuration, $I_0=2.19~\mu\mathrm{A}$, 
$C_g=250~\mathrm{fF}$, and $\Phi_{\mathrm{ext}}=0.5\Phi_0$. 
Further details on the device architecture and experimental implementation 
are reported in~\cite{ranadive_kerr}. 
(a) Two-dimensional maps of the signal gain, expressed as $|S_{21}|$ in dB 
and averaged over the 5--9~GHz signal band, as a function of pump frequency 
and pump power, obtained from simulations (left) and measurements (right). 
The markers indicate the pump conditions considered for the 
frequency-resolved comparison in (b). 
(b) Simulated (red) and measured (blue) signal gain, expressed as
$|S_{21}|$ in dB, as a function of signal frequency at
$f_p=6.5~\mathrm{GHz}$. The lower traces correspond to the pump-off
response, while the upper traces show the development of the gain profile
under pumping. The simulated pumped responses correspond to
$P_p=-76.0$, $-71.9$, and $-70.0~\mathrm{dBm}$, with the two lower-power traces shown with reduced opacity relative to the highest-power trace. The measured pumped response is
reported for an estimated device-input pump power of
$P_p=-67~\mathrm{dBm}$.}
\label{fig:sim_vs_meas}
\end{figure}

The model is benchmarked against the experimental characterization of
the device. Fig.~\ref{fig:sim_vs_meas} compares the measured and
simulated gain profiles for different pump frequencies and pump powers
at the same device configuration. As shown in
Fig.~\ref{fig:sim_vs_meas}(b), the nonlinear harmonic balance
simulations reproduce the main experimental features, including the
evolution of the gain profile with the pump conditions. 
These results support the use of the model for the subsequent
analysis of the linear and pumped circuit responses.

\subsection{Linear metric and configuration selection}

Using this model, we investigate a candidate observable
for constructing the feedback correction term introduced in
Section~\ref{sec:method}. The analysis varies two parameters of the lumped-element
model: the critical current $I_0$ of the large junctions and the
ground capacitance $C_g$, which directly affect the transmission line
response.
As a baseline for quantifying these changes, each circuit configuration
is evaluated through a metric based on the input reflection coefficient
$S_{11}$.
The
metric is defined as
\begin{equation}
\mathcal{M}=
\begin{cases}
\displaystyle
\left\langle S_{11,\mathrm{dB}}(f)\right\rangle_{5\text{--}9\,\mathrm{GHz}},
& S_{11,\mathrm{dB}}(3f_p)>-0.05~\mathrm{dB},\\[5pt]
\mathcal{M}_{\mathrm{penalty}},
& \text{otherwise}.
\end{cases}
\label{eq:metric}
\end{equation}
where
$S_{11,\mathrm{dB}}(f)=20\log_{10}\!\left|S_{11}(f)\right|$ and $f_p$ the target pump frequency.
The first term is the average input reflection within the 5--9 GHz
signal band, with lower values corresponding to reduced reflection
and thus indicating improved impedance matching.
The condition on
$S_{11,\mathrm{dB}}(3f_p)$ provides a linear estimate of third-harmonic suppression: values close to \(0\) dB indicate strong reflection at the third-harmonic frequency, limiting its propagation along the line \cite{levochkina_harmonics}.
Configurations that do not satisfy this condition are excluded by
assigning the penalty value $\mathcal{M}_{\mathrm{penalty}}$.

For each circuit configuration, the metric is evaluated from the linear response as
\begin{equation}
\mathcal{M}_{\mathrm{lin}}(I_0,C_g)
=
\mathcal{M}\!\left[
S_{11}^{\mathrm{lin}}(I_0,C_g)
\right].
\end{equation}
The initial circuit configuration is then selected entirely from the
linear exploration by minimizing the metric:
\begin{equation}
    \left(I_0^{*},C_g^{*}\right)
    =
    \underset{I_0,C_g}{\operatorname{arg\,min}}\,
    \mathcal{M}_{\mathrm{lin}}(I_0,C_g).
\end{equation}

This is the configuration returned by the original single-cycle JCO
workflow before accounting for pump-induced modifications.

\subsection{Pointwise linear-to-nonlinear discrepancy}

In the proposed feedback workflow, the nonlinear simulation is not
repeated over the complete device-parameter space. Instead, it is
performed only for the configuration selected by the linear stage.
The corresponding metric at the selected configuration
$\left(I_0^{*},C_g^{*}\right)$, evaluated from the pumped response at the selected operating point, is
\begin{equation}
\mathcal{M}_{\mathrm{nonlin}}^{*}
=
\mathcal{M}\!\left[
S_{11}^{\mathrm{nonlin}}
\left(I_0^{*},C_g^{*}\right)
\right],
\end{equation}
and its departure from the linear prediction at that point is quantified by
\begin{equation}
    \Delta\mathcal{M}^{*}
    =
    \mathcal{M}_{\mathrm{nonlin}}^{*}
    -
    \mathcal{M}_{\mathrm{lin}}
    \left(I_0^{*},C_g^{*}\right).
\end{equation}

The pointwise linear-to-nonlinear discrepancy
$\Delta\mathcal{M}^{*}$ quantifies the modification of the metric
under pumped operation and is investigated as a candidate input
for constructing the feedback correction term $\mathcal{C}$.

\begin{figure}[!t]
    \centering
    \includegraphics[width=0.95\columnwidth]{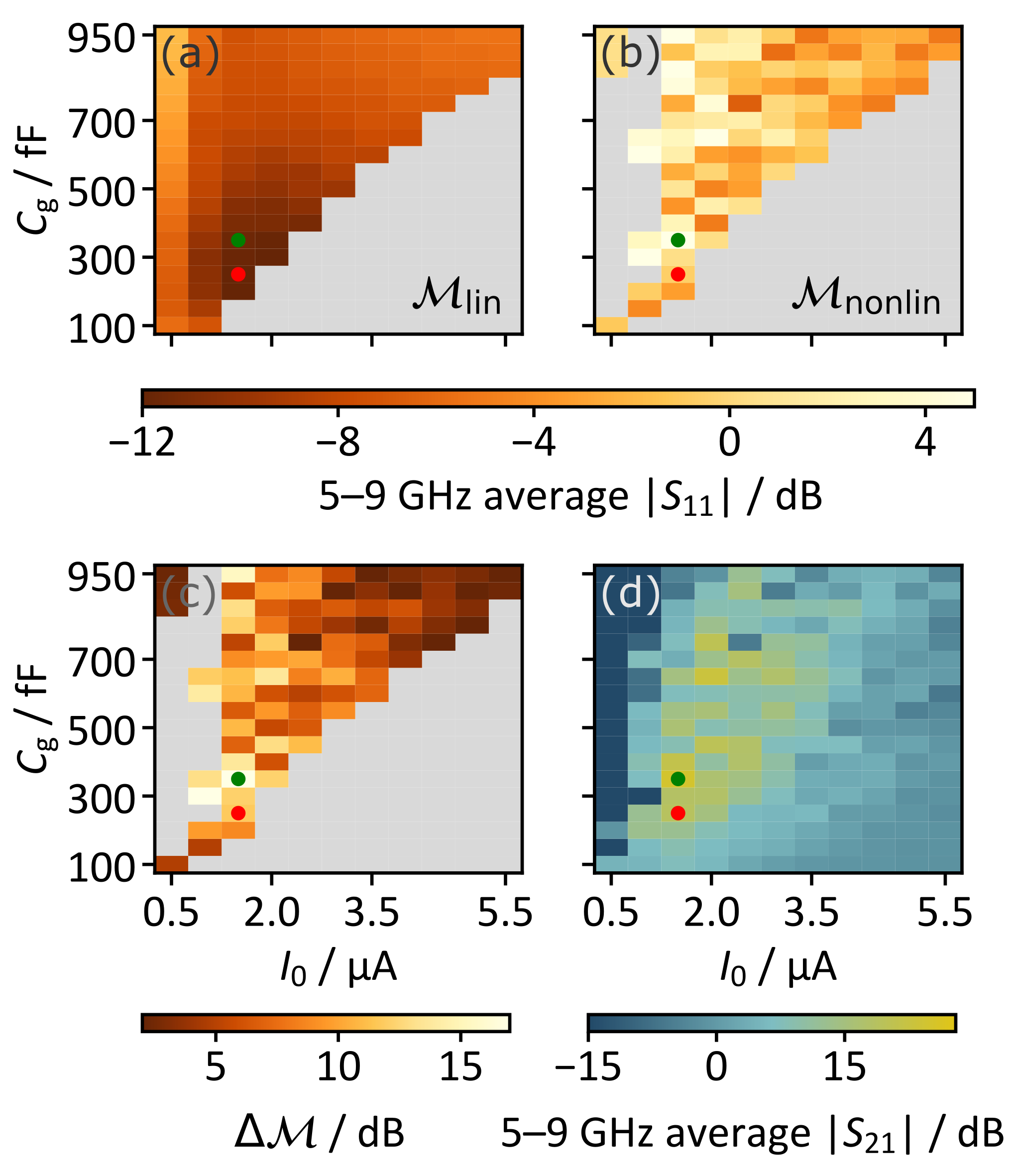}
    \caption{Comparison between the linear and nonlinear circuit responses
over the $(I_0,C_g)$ design space. The flux bias is fixed at
$\Phi_{\mathrm{ext}}=0.5\Phi_0$, and the nonlinear simulations are
performed with a pump at $f_p=7~\mathrm{GHz}$ and
$P_p=-70~\mathrm{dBm}$,
corresponding to the maximum-gain operating point identified for the
configuration selected by the linear optimization.
(a) Linear metric $\mathcal{M}_{\mathrm{lin}}$ derived from the input
reflection coefficient.
(b) Corresponding metric $\mathcal{M}_{\mathrm{nonlin}}$ evaluated
from nonlinear harmonic balance simulations.
(c) Linear-to-nonlinear discrepancy
$\Delta\mathcal{M}
=\mathcal{M}_{\mathrm{nonlin}}-\mathcal{M}_{\mathrm{lin}}$.
(d) Average gain in the 5--9~GHz signal band.
Gray regions indicate configurations rejected by the third-harmonic
condition defined in Eq.~\eqref{eq:metric}, for which the metric is
replaced by the penalty value and the corresponding configurations
are excluded from the optimization.
The red marker indicates the configuration selected by the linear
optimization,
$\left(I_0^{*},C_g^{*}\right)
=\left(1.5~\mu\mathrm{A},250~\mathrm{fF}\right)$,
while the green marker indicates the maximum average gain
configuration,
$\left(1.5~\mu\mathrm{A},350~\mathrm{fF}\right)$.}
    \label{fig:ref_gain}
\end{figure}

To assess the relevance of this correction, the pumped response of the metric $\mathcal{M}_{\mathrm{nonlin}}(I_0,C_g)$ and the discrepancy function $\Delta\mathcal{M}(I_0,C_g)=\mathcal{M}_{\mathrm{nonlin}}(I_0,C_g)-\mathcal{M}_{\mathrm{lin}}(I_0,C_g)$ are
also evaluated over the complete parameter space. This
exhaustive sweep is used only as a diagnostic and is not
required by the proposed feedback workflow.
Fig.~\ref{fig:ref_gain} compares (a) the linear metric
$\mathcal{M}_{\mathrm{lin}}$, (b) the corresponding pumped response
metric $\mathcal{M}_{\mathrm{nonlin}}$, (c) the discrepancy
$\Delta\mathcal{M}$, and (d) the gain over the explored design space.

As shown in Fig.~\ref{fig:ref_gain}(c), $\Delta\mathcal{M}$ exhibits a
pronounced dependence on the circuit configuration, 
showing that the
pumped response does not result in a uniform modification of the linear
metric.
Qualitatively, the region containing the highest-gain
configuration in Fig.~\ref{fig:ref_gain}(d) is also characterized by a pronounced modification of
the metric.
In particular, the configuration selected by the linear metric, marked
in red in Fig.~\ref{fig:ref_gain}, is
$(I_0,C_g)=(1.5~\mu\mathrm{A},250~\mathrm{fF})$ and gives
$\Delta\mathcal{M}\simeq11.73~\mathrm{dB}$ with an average gain of
$18.78~\mathrm{dB}$.
The maximum-gain configuration, marked in green in
Fig.~\ref{fig:ref_gain}, is instead
$(I_0,C_g)=(1.5~\mu\mathrm{A},350~\mathrm{fF})$, with
$\Delta\mathcal{M}\simeq17.37~\mathrm{dB}$ and an average gain of
$24.89~\mathrm{dB}$.

These observations suggest that the variation of the reflection metric contains information related to the amplifier performance that is not captured by the linear response alone. The shift of the maximum-gain configuration along the $C_g$ direction further suggests that the configuration selected by the linear stage can be refined using this pumped response information. These results motivate the investigation of $\Delta\mathcal{M}^{*}$ in subsequent optimization cycles.

\section{Conclusion}
\label{sec:conclusion}

In this work, the JCO workflow was extended with a nonlinear feedback
strategy intended to account for pump-induced effects that are not
captured during the initial linear optimization. The approach was
applied to a 4WM JTWPA with a Kerr-reversal geometry. 
The model was first validated against experimental gain
measurements over different pump frequencies and powers, successfully
reproducing the main evolution of the amplifier response.

The validated model was then used to investigate a linear metric based
on the reflection coefficient in the signal band, while excluding
configurations exhibiting insufficient third-harmonic suppression. The
difference between this metric in the linear and pumped regimes
exhibits qualitative features similar to those of the gain
landscape. In particular, the maximum-gain configuration differs from
the linear optimum and lies in a region characterized by a pronounced
metric discrepancy.

These observations motivate the investigation of the pointwise
linear-to-nonlinear metric variation as a candidate input to the
nonlinear correction operator. In the proposed workflow, this quantity
can be obtained from the nonlinear simulation of only the configuration
selected by the linear stage, preserving the computational advantage of
predominantly linear parameter-space exploration. Future work will
quantitatively assess the relation between the metric discrepancy and
the amplifier performance, define the corresponding update rule, potentially incorporating terms related to the device dispersion, and
evaluate its convergence and effectiveness over successive optimization
cycles.
The data and code used to generate the results presented in this work are available on Zenodo \cite{palumbo_zenodo}.

\section*{Acknowledgment}
The authors gratefully acknowledge Silent Waves for providing the Josephson Traveling-Wave Parametric Amplifier used in this work.

\end{document}